\documentclass[10pt,journal,comsoc]{IEEEtran}

\usepackage{amsmath}
\usepackage{cite}
\usepackage{graphicx}
\usepackage{amsfonts}
\usepackage{latexsym}
\usepackage{subfigure}
\usepackage{algorithm}
\usepackage{algorithmic}
\usepackage{color}
\usepackage{times}
\usepackage{epsfig, graphics}
\usepackage{bm}
\usepackage{subfigure}
\usepackage{graphicx,wrapfig,lipsum}
\usepackage{longtable}

\begin{document}

\title{Adversarial Machine Learning for 5G Communications Security
\thanks{Y. E. Sagduyu, T. Erpek and Y. Shi are with Intelligent Automation, Inc., Rockville, MD, USA; T. Erpek and Y. Shi are also with Virginia Tech, Arlington, VA, USA and Blacksburg, VA, USA, respectively. Email: \{ysagduyu, terpek, yshi\}@i-a-i.com.
}
\thanks{This effort is supported by the U.S. Army Research Office under contract W911NF-17-C-0090. The content of the information does not necessarily reflect the position or the policy of the U.S. Government, and no official endorsement should be inferred.}
}
\author{Yalin E. Sagduyu, Tugba Erpek, and Yi Shi}

\maketitle

\begin{abstract}
Machine learning provides automated means to capture complex dynamics of wireless spectrum and support better understanding of spectrum resources and their efficient utilization. As communication systems become smarter with cognitive radio capabilities empowered by machine learning to perform critical tasks such as spectrum awareness and spectrum sharing, they also become susceptible to new vulnerabilities due to the attacks that target the machine learning applications. This paper identifies the emerging attack surface of adversarial machine learning and corresponding attacks launched against wireless communications in the context of 5G systems. The focus is on attacks against (i) spectrum sharing of 5G communications with incumbent users such as in the Citizens Broadband Radio Service (CBRS) band and (ii)  physical layer authentication of 5G User Equipment (UE) to support network slicing. For the first attack, the adversary transmits during data transmission or spectrum sensing periods to manipulate the signal-level inputs to the deep learning classifier that is deployed at the Environmental Sensing Capability (ESC) to support the 5G system. For the second attack, the adversary spoofs wireless signals with the generative adversarial network (GAN) to infiltrate the physical layer authentication mechanism based on a deep learning classifier that is deployed at the 5G base station. Results indicate major vulnerabilities of 5G systems to adversarial machine learning. To sustain the 5G system operations in the presence of adversaries, a defense mechanism is presented to increase the uncertainty of the adversary in training the surrogate model used for launching its subsequent attacks.
\end{abstract}

\begin{IEEEkeywords}
Adversarial machine learning, deep learning, 5G, spectrum sharing, authentication, network slicing, GAN, jamming, spoofing
\end{IEEEkeywords}

\section{Introduction}

As the fifth generation mobile communications technology, 5G supports emerging applications such as smart warehouses, vehicular networks, virtual reality (VR) and augmented reality (AR) with unprecedented rates enabled by recent advances in massive MIMO, mmWave communications, network slicing, small cells, and internet of things (IoT). Complex structures of waveforms, channels, and resources in 5G cannot be reliably captured by simplified analytical models driven by expert knowledge. As a data-driven approach, \emph{machine learning} has emerged as a viable alternative to support 5G communications by learning from and adapting to the underlying spectrum dynamics \cite{Jiang2016}. Empowered by recent advances in algorithmic design and computational hardware resources, \emph{deep learning} shows strong potential to learn the high-dimensional data characteristics of wireless communications beyond conventional machine learning techniques \cite{Erpek19} and offers novel solutions to critical tasks of detection, classification, and prediction in 5G systems. 

As machine learning becomes a core part of next-generation communication systems, there is an increasing concern about the vulnerability of machine learning to adversarial effects. To that end, smart adversaries may leverage emerging machine learning techniques to infer vulnerabilities in 5G systems and tamper with the learning process embedded in 5G communications. The problem of learning in the presence of adversaries is the subject to the study of \emph{adversarial machine learning} that has received increasing attention in computer vision and natural language processing (NLP) domains \cite{AML1, AMLbook, Sec2}. Due to the shared and open nature of wireless medium, wireless applications are highly susceptible to adversaries such as jammers and eavesdroppers that can manipulate the training and testing processes of machine learning over the air. While there is a growing interest in designing attacks on machine learning-driven data and control planes of wireless communications \cite{ErpekTCCN, Sagduyu19, SagduyuWhenWireless, Adesina2020}, adversarial machine learning has not been considered yet for sophisticated communication systems such as 5G.  

5G systems are designed to operate in frequency bands from 450 MHz to 6 GHz, and 24.250 GHz to 52.600 GHz (millimeter-wave bands) including the unlicensed spectrum. While some of these bands are dedicated to commercial use of 5G, some other ones are opened for \emph{spectrum co-existence} of 5G with other legacy wireless systems. In particular, the U.S. Federal Communications Commission (FCC) has adopted rules for the \emph{Citizens Broadband Radio Service (CBRS) band} to allow the use of commercial communications systems in the 3550-3700 MHz band in an opportunistic manner by treating CBRS users as incumbent. These communications systems including 5G systems are required to vacate the band once the ranging (radar) signal is detected by the Environmental Sensing Capability (ESC) system as an incumbent CBRS user \cite{FCCCBRS}. Radar signal detection and classification is a complex problem considering the unpredictable utilization patterns, channel effects and interference from other commercial systems operating in the same band and interference leakage from radar systems operating in different bands. Deep neural networks can capture complex spectrum effects in these bands and perform superior performance compared to conventional signal detection and classification techniques \cite{Souryal2019, LeesTCCN}. On the other hand, the use of deep neural networks may expose the system to adversarial attacks, as adversaries can tamper with both data and control plane communications with smart jamming and consequently prevent efficient spectrum sharing of 5G with incumbent users.   

Another aspect of 5G that is vulnerable to adversarial machine learning is \emph{network slicing}. As an emerging concept that enables 5G to serve diverse applications (such as IoT and autonomous driving) with different performance requirements (such as throughput and latency) on heterogeneous platforms, network slicing multiplexes virtualized and independent logical networks on a common physical network infrastructure \cite{ZhangTWC}. Our focus is on network slicing in the 5G radio access network (RAN) \cite{IAINetSlicing}. In this setting, both the network slice manager and the user equipment (UE) cannot be trusted in general \cite{NGMNAlliance} and smart adversaries may impersonate their roles. As part of physical layer security, deep learning can be used by 5G for RF fingerprinting to classify and authenticate signals received at network slice manager or host machines. However, by exploring the vulnerabilities of \emph{physical layer signal authentication}, adversarial machine learning provides novel techniques to spoof wireless signals that cannot be reliably distinguished from intended users even when deep neural networks are utilized for wireless signal classification. 

Compared to other data domains, there are several unique challenges when we apply adversarial attacks in wireless communications.
\begin{enumerate}
\item  In data domains such as computer vision and NLP, it may be possible that the input to the machine learning algorithm (e.g., a target classifier) is directly manipulated by an adversary, e.g., by querying an online application programming interface (API). However, an adversary in the wireless domain cannot directly collect the same input data (e.g., spectrum sensing data) as the communication system's transmitter, due to the different channel and interference conditions perceived by the target model and the adversary. As a result, the features (input) to the machine learning algorithm are different for the same instance. 
\item The adversary in a wireless domain cannot directly obtain the output (label) of the target machine learning algorithm, since the output is used by the target system such as 5G only and thus is not available to any other wireless node outside the network. As a result, the adversary needs to observe the spectrum to make sense of the outputs of the target machine learning algorithm but cannot necessarily collect exactly the same outputs (e.g., classification labels). 
\item The adversary in a wireless domain cannot directly manipulate the input data to a machine learning algorithm. Instead, it can only add its own transmissions on top of existing transmissions (if any) over the air (i.e., through channel effects) to change the input data (such as spectrum sensing data) indirectly. 
\item Input features of the machine learning algorithm to be used by the adversary may differ from those used by the target communication system depending on the differences in the underlying waveform and receiver hardware characteristics at the adversary and the communication system. 
\end{enumerate}

By accounting for these challenges, we will describe how to apply adversarial machine learning to the 5G communication setting. In particular, we will discuss the \emph{vulnerabilities of machine learning applications in 5G} with two motivating examples and show how adversarial machine learning provides a new attack surface in 5G communication systems:
\begin{enumerate}
    \item \textit{Attack on spectrum sharing of 5G with incumbent users such as in CBRS}: The ESC system senses the spectrum and uses a deep learning model to detect the radar signals as an incumbent user. If the incumbent user is not detected in a channel of interest, the 5G transmitter, in our case the 5G base station, namely the 5G gNodeB, is informed and starts with communications to the 5G UE. Otherwise, the 5G gNodeB cannot use this channel and the Spectrum Access system (SAS) reconfigures the 5G system's spectrum access (such as vacating this particular channel) to avoid interference with the incumbent signals. By monitoring spectrum dynamics of channel access, the adversary builds a surrogate model of the deep neural network architecture based on its sensing results, predicts when a successful 5G transmission will occur, and jams the communication signal accordingly in these predicted time instances. We consider jamming both data transmission and spectrum sensing periods. The latter case is stealthier (i.e., more difficult to detect) and more energy efficient as it only involves jamming of the shorter period of spectrum sensing before data transmission starts and forces the 5G gNodeB into making wrong transmit decisions. We show that this attack reduces the 5G communication throughput significantly. 
    \item \textit{Attack on network slicing}: The adversary transmits spoofing signals that mimic the signal characteristics of the 5G UE when requesting a network slice from the host machine. This attack potentially allows the adversary to starve resources in network slices after infiltrating through the authentication system built at the 5G gNodeB. A generative adversarial network (GAN) \cite{GoodfellowGAN14} is deployed at an adversary pair of a transmitter and a receiver that implement the generator and discriminator, respectively, to generate synthetic wireless signals that match the characteristics of a legitimate UE's 5G signals that the 5G gNodeB would expect to receive. We show that this attack allows an adversary to infiltrate the physical layer authentication of 5G with high success rate. 
\end{enumerate}

These novel attacks leave a smaller footprint and are more energy-efficient compared to conventional attacks such as jamming of data transmissions. As a countermeasure, we present a \emph{proactive defense} approach to reduce the performance of the inference attack that is launched as the initial step to build an adversarial (surrogate) model that captures wireless communication characteristics. Other attacks are built upon this model. The 5G system as the defender makes the adversarial model less accurate by deliberately making a small number of wrong decisions (such as in spectrum sensing, data transmission, or signal authentication). The defense carefully selects which decisions to flip with the goal of maximizing the uncertainty of the adversary while balancing the impact of these controlled decision errors on its own performance.

This rest of the paper is organized as follows. We introduce adversarial machine learning and describe the corresponding attacks in Section \ref{subsec:AML}. Then, we extend the use of adversarial machine learning to the wireless communications and discuss the domain-specific challenges in Section \ref{AMLforWireless}. After identifying key vulnerabilities of machine learning-empowered 5G solutions, Section \ref{AMLfor5G} introduces two attacks built upon adversarial machine learning against 5G communications and presents a defense mechanism. Section \ref{Conclusion} concludes the paper.  

\section{Adversarial Machine Learning} \label{subsec:AML}
While there is a growing interest in applying machine learning to different data domains and deploying machine learning algorithms in real systems, it has become imperative to understand vulnerabilities of machine learning in the presence of adversaries. To that end, adversarial machine learning \cite{AML1, AMLbook,Sec2} has emerged as a critical field to enable safe adoption of machine learning subject to adversarial effects. One example that has attracted recent attention involves machine learning applications offered to public or paid subscribers via APIs; e.g., Google Cloud Vision \cite{GoogleCloud} provides cloud-based machine learning tools to build machine learning models. This online service paradigm creates security concerns of adversarial inputs to different machine learning algorithms ranging from computer vision to NLP \cite{ShiISSPIT18, ShiHST2018}. As another application domain, automatic speech recognition and voice controllable systems were studied in terms of the vulnerabilities of their underlying machine learning algorithms \cite{Carlini16,Zhang17}. As an effort to identify vulnerabilities in autonomous driving, attacks on self-driving vehicles were demonstrated in \cite{Kurakin16AtScale}, where the adversary manipulated traffic signs to confuse the learning model.

\begin{figure} 
	\centering
	\fbox{\includegraphics[width=0.45\textwidth]{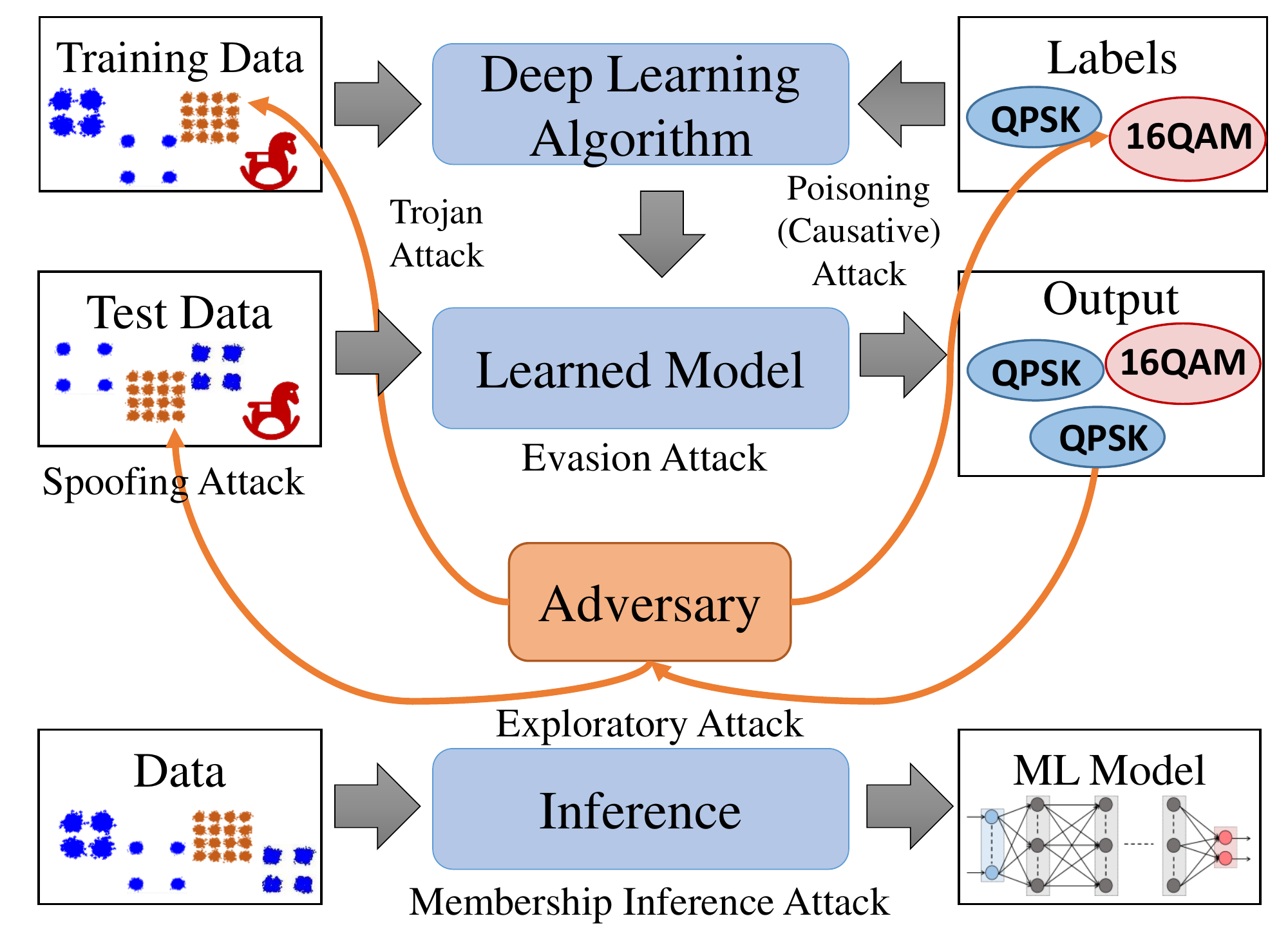}}
	\caption{Taxonomy of attacks built upon adversarial machine learning.}
	\label{fig:MLTaxonomy}
\end{figure}

The manipulation in adversarial machine learning may happen during the training or inference (test) time, or both. During the training time, the goal of the adversary is to provide wrong inputs (features and/or labels) to the training data such that the machine learning algorithm is not properly trained. During the test time, the goal of the adversary is to provide wrong inputs (features) to the machine algorithm such that it returns wrong outputs. As illustrated in Fig.~\ref{fig:MLTaxonomy}, attacks built upon adversarial machine learning can be categorized as follows.  

\begin{enumerate}
\item \textit{Attack during the test time.}
\begin{enumerate}
    \item \textit{Inference (exploratory) attack}: The adversary aims to infer the machine learning architecture of the target system to build a shadow or surrogate model that has the same functionality as the original machine learning architecture \cite{Sec2, Barreno, Tramer, Papernot, ShiHST17, WuCSF16}. This corresponds to a white-box or black-box  attack depending on whether the machine learning model such as the deep neural network structure is available to the adversary, or not. For a black-box attack, the adversary queries the target classifier with a number of samples and records the labels. Then, it uses this labeled data as its own training data to train a functionally-equivalent (i.e., statistically similar) deep learning classifier, namely a surrogate model. Once the machine learning functionality is learned, the adversary can use the inference results obtained from the surrogate model for subsequent attacks such as confidence reduction or targeted misclassification.
    
    \item \emph{Membership inference attack}: The adversary aims to determine if a given data sample is a member of the training data, i.e., if a given data sample has been used to train the machine learning algorithm of interest \cite{Shokri18, Leino19, Mittal19, Gong19}. Membership inference attack is based on the analysis of overfitting to check whether a machine learning algorithm is trained for a particular data type, e.g., a particular type of images. By knowing which type of data the machine learning algorithm is trained to classify, the adversary can then design a subsequent attack more successfully.
    
    \item \textit{Evasion attack}: The adversary manipulates test data of a machine learning algorithm by adding carefully crafted adversarial perturbations \cite{ShiMILCOM17, Papernot16, Kurakin16, Moosavi15}. Then, the machine learning model runs on erroneous test data and makes errors in test time, e.g., a classifier is fooled into accepting an adversary as legitimate. The samples with output labels that are closer to the decision region can be used by the adversary to increase the probability of error at the target machine learning algorithm. 
    
    \item \textit{Spoofing attack}: The adversary generates synthetic data samples from scratch rather than adding perturbations to the real ones. The GAN can be used for data spoofing by generating a synthetic dataset that is statistically similar to the original dataset \cite{ShiWiseML,ShiSpoofingTCCN, DavasliogluICC}. The GAN consists of two deep neural networks, one acting as a generator and the other one acting as a discriminator. The generator generates spoofing signals and the discriminator aims to detect whether the received signal is spoofed, or not. Then, the generator and the discriminator play a mini-max game to optimize their individual performance iteratively in response to each other's actions. After they converge, the generator's deep neural network is used to generate spoofing signals. 
    
    \end{enumerate}
 \item \textit{Attack during the training time.} \\
    \textit{Poisoning (causative) attack}: The adversary manipulates the training process by either directly providing wrong training data or injecting perturbations to the training data such that the machine learning model is trained with erroneous features and thus it makes errors later in test time \cite{ShiMILCOM17, Pi16, Alfeld16}. This attack is stealthier than the evasion attack (as the training period is typically shorter than the test period). To select which training data samples to tamper with, the adversary first runs samples through the inferred surrogate model and then changes their labels and sends these mislabeled samples as training data to the target classifier provided that their deep learning scores are far away from the decision region of the surrogate model.
    
    \item \textit{Attack during both training and test times.} \\
    \textit{Trojan (backdoor or trapdoor) attack}: The adversary slightly manipulates the training data by inserting Trojans, i.e., triggers, to only few training data samples by modifying some data characteristics (e.g., putting stickers on traffic signs) and changing the labels of these samples to a target label (e.g., from the stop sign to the speed limit sign). This poisoned training data may be used to train the machine learning model. In test time, the adversary feeds the target classifier with input samples embedded with the same characteristics that were added as triggers by the adversary during training. The goal of the adversary is to cause errors when machine learning is run on data samples poisoned with triggers. In the meantime, clean (unpoisoned) samples without triggers should be processed correctly. Since only few samples of Trojans are inserted, this attack is harder to detect than both evasion and causative attacks. The disadvantage of the Trojan attack is that it needs to be launched during both training and test times, i.e., the adversary needs to have access to and manipulate both training and test data samples.
\end{enumerate}

Various defense mechanisms have been developed in the literature against adversarial machine learning attacks in computer vision, NLP, and other data domains. The core of the defense is to make the machine learning algorithm robust to the anticipated attacks. One approach against evasion attacks is randomized smoothing during training, where a number of small Gaussian noise samples are added to each training data sample to augment the training dataset. Then, the classifier that is trained with this augmented training dataset becomes robust against adversarial inputs in test time \cite{certified}. The defense can be further certified by adding perturbations to training data and generating a certificate to bound the expected error due to perturbations added later in test time \cite{certified}. Note that this defense assumes that the attack provides erroneous input by adding perturbations to test data. We consider a proactive defense mechanism in Section~\ref{subsec:defense} against adversarial machine learning in wireless communications. The goal of this defense is to provide a small number of carefully crafted wrong inputs to the adversary, as it builds (trains) its attack scheme, and thus prevent the adversary from building a high-fidelity surrogate model.    

\section{Adversarial Machine Learning in \\  Wireless Communications} \label{AMLforWireless}
Machine learning finds rich applications in wireless communications, including spectrum access \cite{ShiDyspan}, signal classification \cite{SoltaniMilcom}, beamforming \cite{Mismar}, beam selection \cite{Tarun}, channel estimation \cite{GLi2019}, channel decoding \cite{HoydisCISS}, physical layer authentication \cite{DeepWifi}, and transmitter-receiver scheduling \cite{NofMilcom}. In the meantime, there is a growing interest in bridging machine learning and wireless security in the context of adversarial machine learning \cite{ErpekTCCN, Sagduyu19}. In Section~\ref{subsec:attackSch}, we  discuss how different attacks in adversarial machine learning presented  in Section \ref{subsec:AML} can be adapted to the wireless communications. Then, we identify the domain-specific challenges on applying adversarial machine learning to wireless communications in Section \ref{subsec:challenge}. Finally, we discuss the state-of-the-art defense techniques against the adversarial attacks in wireless communications in Section \ref{subsec:defense}.    

\subsection{Wireless Attacks built upon Adversarial Machine Learning} \label{subsec:attackSch}
\begin{figure*} 
	\centering
	\fbox{\includegraphics[width=0.8\textwidth]{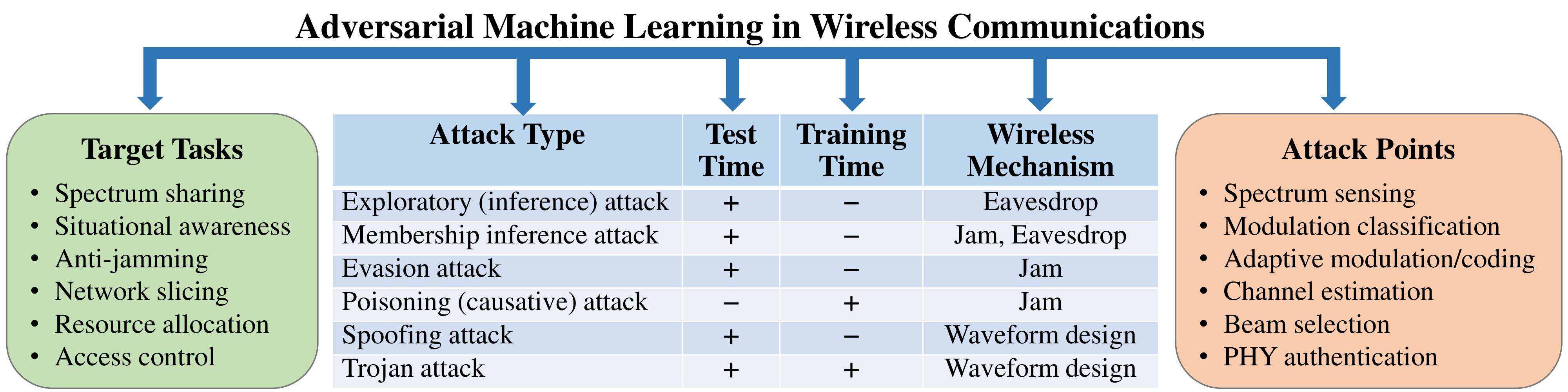}}
	\caption{Key points of adversarial machine learning in wireless communications.
	\label{fig:AMLwireless}}
\end{figure*}

Fig.~\ref{fig:AMLwireless} illustrates target tasks, attack types, and attack points of adversarial machine learning when applied to the wireless communications. As the motivating scenario to describe different attacks, we consider a canonical wireless communication system with one transmitter $T$, one receiver $R$, and one adversary $A$. This setting is instrumental in studying conventional jamming \cite{conventional1, conventional2, conventional3} and defense strategies in wireless access, and can be easily extended to a network scenario with multiple transmitters and receivers \cite{ErpekTCCN}. The communication system shares the spectrum with a background traffic source $B$, whose transmission behavior is not known by $T$ and $A$. 

In wireless domains, the following attacks have been considered against a machine learning-based classifier.

\begin{itemize}
\item \emph{Exploratory (inference) attack}. Transmitter $T$ uses a machine learning-based classifier $C_T$ for a specific task such as spectrum sensing and adversary $A$ aims to build a classifier $C_A$ (namely, the surrogate model) that is functionally the same as (or similar to) the target classifier 
$C_T$ \cite{Erpek19, Shi1805}. The observations and labels at $A$ will differ compared to the ones at $C_T$ since $T$ and $A$ are not co-located and they experience different channels. 

\item \emph{Membership inference attack}. Adversary $A$ aims to determine whether a given data sample is in the training data of $C_T$. One example of this attack in the wireless domain is to identify whether a target classifier is trained against signals from a particular transmitter, or not \cite{MAIShi}.  

\item \emph{Evasion attack}. Adversary $A$ aims to determine or craft input samples that the target classifier $C_T$ cannot reliably classify. Wireless examples of evasion attacks include manipulation of inputs to the spectrum sensing \cite{Sagduyu19, Shi1810, Sagduyu1906}, modulation classification \cite{Larsson2018, Headley2019, Deniz2019, Deniz2019-2, Flowers2019, Silvija2019, Silvija2019b, Kim2020, Kim2020-2, Kim2020-4, Kim2020-5}, channel access \cite{Gursoy1, Gursoy2}, autoencoder-based physical layer design, \cite{Sadeghi19}, and eavesdroppers for private communications \cite{Kim2020-3}. 
    
\item \emph{Poisoning (Causative) attack}. Adversary $A$ provides falsified samples to train or retrain the target classifier $C_T$ such that $C_T$ is not properly trained and makes significantly more errors than usual in test time. Poisoning attacks were considered against spectrum sensing at an individual wireless receiver in \cite{Sagduyu19,Sagduyu1906} and against cooperative spectrum sensing at multiple wireless receivers in \cite{LuoArxiv, LuWiseML}. Note that these poisoning attacks against sensing decisions are extensions of the conventional spectrum sensing data falsification (SSDF) attacks \cite{SSDF} to the adversarial machine learning domain but they are more effective as they manipulate the training process more systematically for the case of machine learning-based spectrum sensors.  

\item \textit{Spoofing attack}: Adversary $A$ generates spoofing signals that impersonate transmissions originated from $T$. The GAN can be used to generate synthetic forms of wireless signals that can be used not only to augment training data (e.g., for a spectrum sensing classifier \cite{DavasliogluICC}) but also to fool a target classifier \cite{ShiWiseML, ShiSpoofingTCCN}. To spoof wireless signals in a distributed setting, two adversaries (one transmitter and one receiver) assume roles of a generator and a discriminator of the GAN to spoof and discriminate signals, respectively, with two separate deep neural networks.

\item \emph{Trojan attack}. Adversary $A$ provides falsified samples to train the target classifier $C_T$ such that $C_T$ works well in general but provides incorrect results if a certain trigger is activated. For example, small phase shifts can be added as triggers to wireless signals (without changing the signal amplitude) to launch a stealthy Trojan attack against a wireless signals classifier \cite{Davaslioglu19}. 
\end{itemize}

\subsection{Domain-specific Challenges for Adversarial Machine Learning in Wireless Communications} \label{subsec:challenge}
Wireless applications of adversarial machine learning are different from other data domains such as computer vision in four main aspects. 
\begin{enumerate}
\item The adversary and the defender may not share the same features (such as received signals) as channels and interference effects observed by them are different. 
\item The adversary and the defender may not share the same labels (i.e., machine learning outputs). For example, the defender may aim to classify channel as busy or not during spectrum sensing, whereas the adversary may need to decide on whether the defender will have a successful transmission or not. These two objectives may differ due to different channel and interference effects observed by the adversary and the defender. 
\item The adversary may not directly manipulate the input data to the machine learning algorithm, as wireless users are typically separated in location and receive their input from wireless signals transmitted over the air. Therefore, it is essential to account for channel effects when designing wireless attacks and quantifying their impact. 
\item The type of data observed by the adversary depends on the underlying waveform and receiver hardware of the adversary. While adversarial machine learning may run on raw data samples such as pixels in computer vision applications, the adversary in a wireless domain may have to use different types of available radio-specific features such as I/Q data and received signal strength indicator (RSSI) that may differ from the features used by the target machine learning system.
\end{enumerate}

With the consideration of the above challenges, we can apply adversarial machine learning to design attacks on wireless communications. In an example of \emph{exploratory} attack, $A$ collects spectrum sensing data and obtains features for its own classifier $C_A$. Unlike the traditional exploratory attack, where $C_A$ provides the same set of prediction results as the target classifier $C_T$, the label that $A$ aims to predict is whether there is a successful transmission from $T$ or not, which can be collected by sensing the acknowledgement message (ACK). For that purpose, $A$ collects both input (features) and output (label) for $C_A$. If for an instance, $A$ can successfully predict whether there will be an ACK, $C_A$ can predict the correct result. Deep learning is successful in building the necessary classifier $C_A$ for $A$ in exploratory attacks \cite{ErpekTCCN}. 

We can further design an example of \emph{evasion attack} as follows. If $A$ predicts a successful transmission, it can either transmit in the sensing phase (to change features to $C_T$) \cite{Sagduyu19, Shi1810} 
or in the data transmission phase (to jam data) \cite{ErpekTCCN}. For the first case, most idle channels are detected as busy by $C_T$ and thus throughput is reduced almost to zero \cite{Sagduyu19}. For the second case, many successful transmissions (if no jamming) are jammed and thus throughput is reduced significantly \cite{Sagduyu1906}.

In a \emph{causative attack}, $C_T$ is updated by additional training data collected over time and $A$ attempts to manipulate this retraining process. For example, if $T$ transmits but does not receive an ACK, the prediction by $C_T$ is incorrect and additional training data is collected. $C_T$ is expected to improve with additional training data. However, $A$ can again either transmit in the sensing phase or in the data transmission phase to manipulate the training process. For the first case, features are changed \cite{Sagduyu19}. For the second case, many transmissions that would be otherwise successful are jammed and consequently their labels are changed \cite{Sagduyu1906}. For both cases, throughput (using updated $C_T$) is reduced significantly.

\emph{Membership inference attack} is based on the analysis of overfitting. Note that features can be either useful (used to predict the channel status) or biased (due to the different distributions of training data and general test data) information.
$C_T$ is optimized to fit on useful and biased information ($F_u$ and $F_b$). Fitting on $F_b$ corresponds to overfitting, which provides correct classification on the given training data but wrong classification on general test data.
In a white box attack, $A$ studies parameters in $C_T$ based on local linear approximation for each layer and the combination of all layers. This approach builds a classifier for membership inference that can leak private information of a transmitter such as waveform, channel, and hardware characteristics by observing spectrum decisions based on the output of a wireless signal classifier \cite{MAIShi}.

In the \emph {Trojan attack}, $A$ slightly manipulates training data by inserting triggers to only few training data samples by modifying their phases and changing the labels of these samples to a target label. This poisoned training data is used to train $C_T$. In test time, $A$ transmits signals with the same phase shift that was added as a trigger during training time. $R$ accurately classifies clean (unpoisoned) signals without triggers, but misclassifies signals poisoned with triggers \cite{Davaslioglu19}.

\subsection{Defense Schemes against Adversarial Machine Learning} \label{subsec:defense}
 The basis of many attacks such as evasion and poisoning attacks discussed in Section~\ref{subsec:attackSch} is the exploratory attack that trains a functionally equivalent classifier $C_A$ as a surrogate model for target classifier $C_T$. Once $C_A$ is built, the adversary can analyze this model to understand the behavior of $C_T$, which paves the way for identifying the weaknesses of $C_T$ and then designing subsequent attacks. Therefore, a defense mechanism is needed to mitigate the exploratory attack. One \emph{proactive defense} mechanism is to add controlled randomness to the target classifier $C_T$ such that it is not easy to launch an exploratory attack. For that purpose, transmitter $T$ can transmit when channel is detected as busy or can remain idle when channel is detected as idle. However, such incorrect decisions will decrease the system performance even without attacks. Thus, the key problem is how to maximize the effect of defense while minimizing the impact on system performance. In particular, our approach is to exploit the likelihood score, namely the confidence level, returned by the machine learning classifier such that $T$ performs defense operations only when the confidence is high, thereby maximizing the utility of each defense operation \cite{Shi1805}. This way, with a few  defense operations, the error probability of exploratory attack can be significantly increased  \cite{Sagduyu19} and subsequent attacks such as the evasion attack in the sensing phase can be further mitigated. The number of defense operations can be adapted in a dynamic way by monitoring the performance over time.

Possible approaches against the membership inference attack include the following two. The first approach aims to make the distribution in training data similar to the general test data. When we apply $C_T$ on some samples, if a sample is classified with high confidence, it is likely that this sample contributes to overfitting in the training data and it is removed to make the training data similar to general test data. The second approach aims to remove the biased information in $F_b$. We can analyze the input to any layer in the deep neural network and identify features that play an important role in $F_b$. Then, we can rebuild $C_T$ on features other than identified ones to remove the impact of overfitting.

Trojan attacks can be detected by identifying potential triggers inserted into training data. Since all malicious samples have a particular trigger, we can apply outlier detection methods such as the one based on Median Absolute Deviation (MAD) or clustering to detect this trigger in the Trojan attack. Once the trigger is detected, any sample with this trigger is discarded or its label is switched to mitigate the attack \cite{Davaslioglu19}.

\section{Adversarial Machine Learning in 5G Communications} \label{AMLfor5G}
We present two scenarios to demonstrate adversarial machine learning-based attacks on 5G systems. The details of the first and second scenario and the performance results are presented in Section \ref{subsec:Scenario1} and Section \ref{subsec:Scenario2}, respectively.

\subsection{Scenario 1 - Adversarial Attack on 5G Spectrum Sharing} \label{subsec:Scenario1}

\subsubsection{Attack Setting} \label{subsubsec:Attack1}
\begin{figure} 
	\centering
	\fbox{\includegraphics[width=0.95\columnwidth]{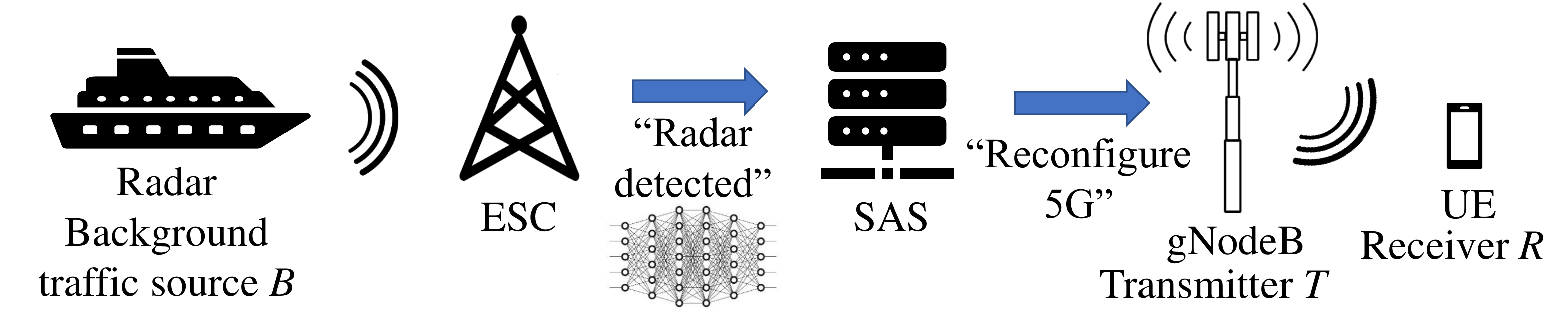}}
	\caption{Scenario 1: Spectrum sharing of 5G with incumbent user (radar) in the CBRS band.
	\label{fig:Scenario1Step1}}
\end{figure}

\begin{figure}[h]
	\centering
	\fbox{\includegraphics[width=0.95\columnwidth]{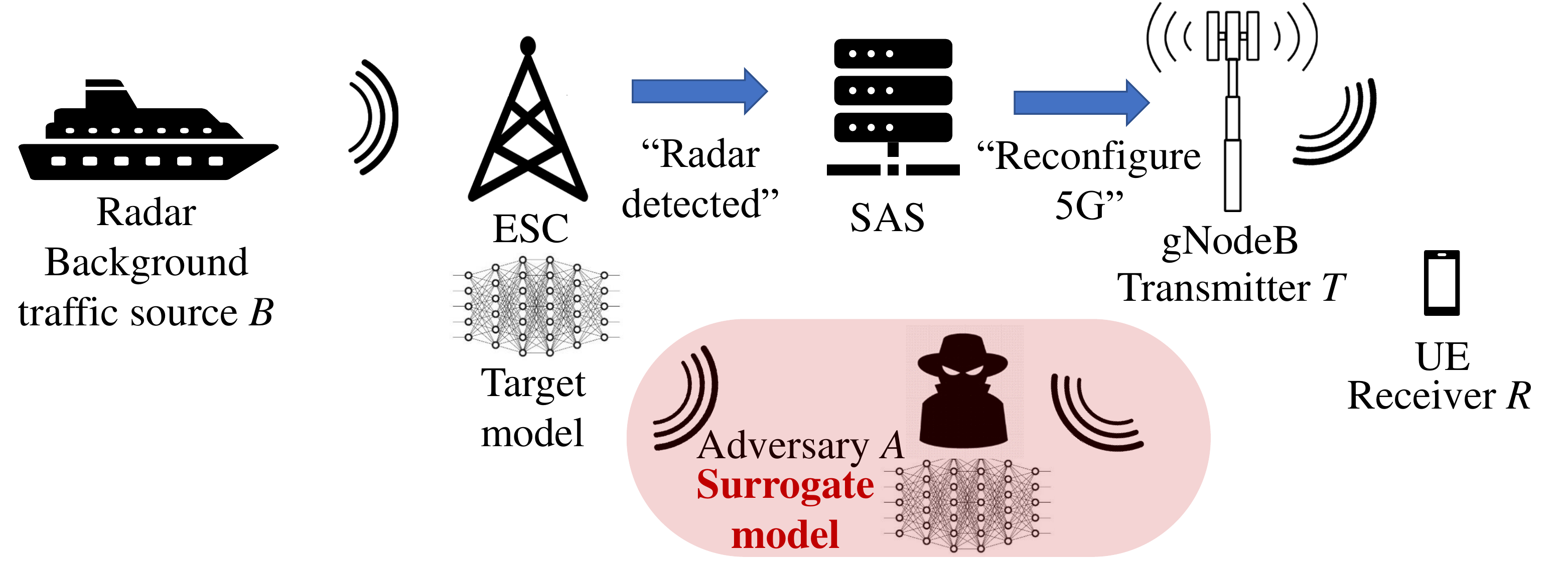}}
	\caption{Scenario 1 - Attack Step 1: Adversary trains adversarial deep learning classifier as the surrogate model to infer the ongoing transmit pattern of the incumbent.
	\label{fig:Scenario1Step2}}
\end{figure}

The operation of 5G communications systems is expected to cover the CBRS band, where 5G users need to share the spectrum with the radar signal. The radar is the incumbent (primary) user of the band and the 5G communications system is the secondary user. 5G transmitter ($T$), namely 5G gNodeB, and receiver ($R$), namely 5G UE, need to communicate when no radar signal is detected in the band as the background traffic source $B$. The ESC system senses the spectrum, decides whether the channel is idle or busy by a machine learning-based classifier $C_T$, and informs its decisions to the SAS. SAS informs $T$ and if the channel is idle, $T$ transmits data. $R$ sends an ACK once it receives data from $T$. This procedure is shown in Fig. \ref{fig:Scenario1Step1}. An adversary $A$ also senses the spectrum and decides when to perform certain attack actions, as shown in Fig. \ref{fig:Scenario1Step2}.

\begin{figure} [h]
	\centering
	\fbox{\includegraphics[width=0.95\columnwidth]{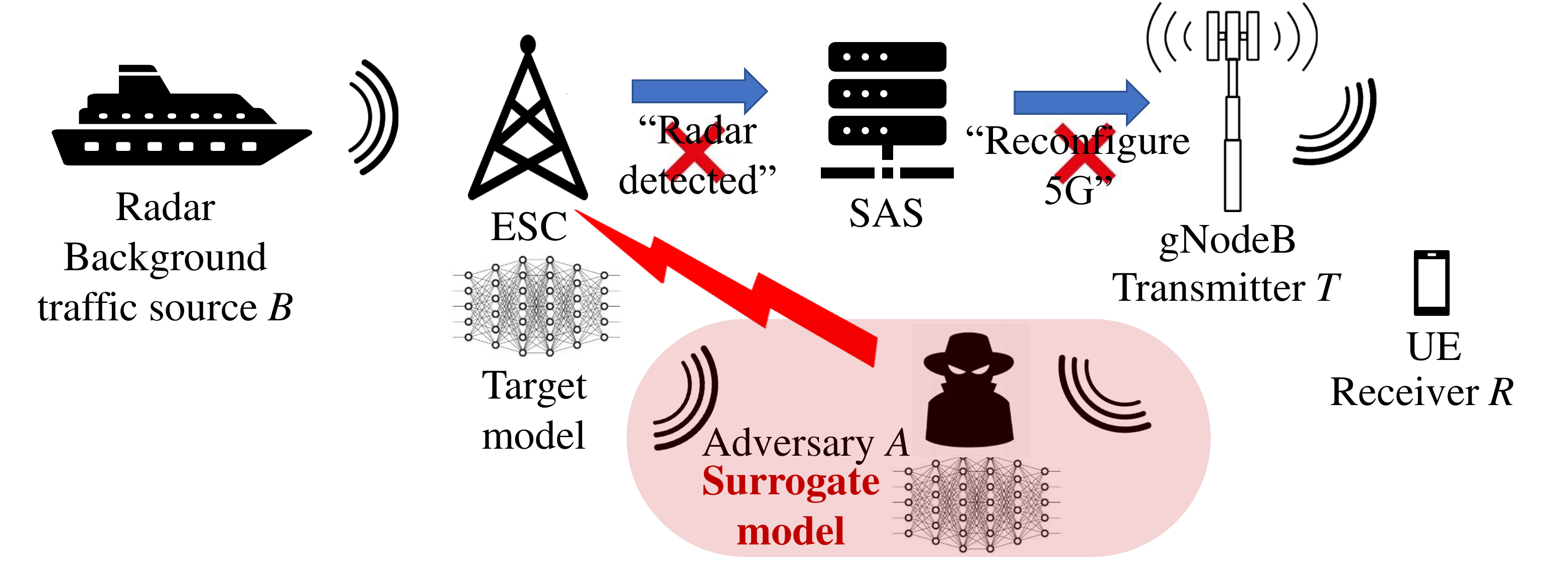}}
	\caption{Scenario 1 - Attack Step 2: Adversary jams the incumbent transmissions by using its surrogate model.
	\label{fig:Scenario1Step3}}
\end{figure}

We consider the following attack actions built upon adversarial machine learning. First, the adversary $A$ trains the surrogate model $C_A$ in the form of an exploratory attack, as shown in Fig. \ref{fig:Scenario1Step2}. Then, $A$ uses the surrogate model to decide when and how to interfere with incumbent user's spectrum access process, as shown in Fig. \ref{fig:Scenario1Step3}. $A$ aims to either jam data transmissions such that $R$ cannot receive data transmission from $T$ or jam the spectrum sensing period such that an idle channel is considered as busy. The first part is a conventional jamming attack and the second part is fooling $T$ into wasting transmit opportunities. The second part corresponds to an evasion attack as it manipulates the sensing inputs into the machine learning algorithm used by $T$ for spectrum access decisions.

Since the adversary $A$ only needs to attack when there will be a successful transmission (if there was no attack), it aims to decide whether there is an ACK returned by the receiver $R$ and uses the presence or absence of ACKs as labels to train its machine learning-based classifier $C_A$.
The process of building $C_A$ is not exactly the same as the conventional exploratory attack (such as the one considered in computer vision applications), where $C_A$ should be the same (or as similar as possible) as $C_T$. Due to different spectrum sensing results at different locations, the inputs to $C_A$ and $C_T$ are different. Further, the output of $C_A$ (i.e., `ACK' or `no ACK') is different than the output of $C_T$ (i.e., `idle' or `busy'). 

\subsubsection{Simulation Setup and Performance Results} \label{subsubsec:Sims}
For this attack, we set up a scenario, where the distance from $T$ to $B$ is $1000$ m  and the distance from $A$ to $B$ is $1010$ m. $T$ builds $C_T$ based on the sensed signal powers, namely the RSSIs. Each data sample consists of $200$ sensing results and $T$ collects $1000$ of those samples. Then, one half of these samples are used for training and the other half are used for testing. We train and test classifiers in TensorFlow and consider the following classifier characteristics.
\begin{itemize}
	\item A feedforward neural network is trained for each classifier with backpropagation algorithm by using the cross-entropy loss function.
	\item Rectified linear unit (ReLU) activation function is used at the hidden layers.
	\item Softmax activation function is used at the output layer.
	\item Batch size is $100$.
	\item Number of training steps is $1000$.
\end{itemize}

The deep neural network structure of classifier $C_T$ is given as follows.
\begin{itemize}
	\item The input layer has $200$ neurons.
	\item The first hidden layer is a dense layer of $512$ neurons. 
	\item The second hidden layer is a dropout layer with dropout ratio of $0.2$.
	\item The third hidden layer is a dense layer of $512$ neurons. 
	\item The fourth hidden layer is a dropout layer with dropout ratio of $0.2$. 
	\item The output layer has two neurons.
\end{itemize}

The monostatic radar signal is simulated in MATLAB as the background signal. Free space model is used to calculate the propagation loss between $B$ and $T$. The classifier $C_T$ has very good performance in the absence of adversaries. It can correctly detect all idle channel instances and most busy channel instances. The error on busy channel detection is $5.6$\%. That is, the 5G system can successfully protect $94.4$\% of radar transmissions while achieving $100$\% throughput (normalized by the best throughput that would be achieved by an ideal algorithm that detects every idle channel correctly). 

\begin{figure} 
	\centering
	\fbox{\includegraphics[width=0.95\columnwidth]{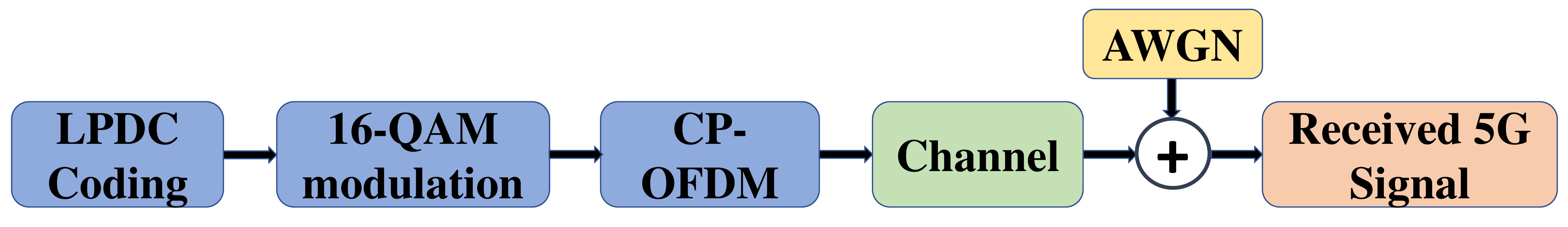}}
	\caption{Generating 5G NR signal at the UE.
	\label{fig:5GCommChain}}
\end{figure}

The 5G communications system in this scenario uses 5G New Radio (NR) signal. The 5G NR signal is generated using MATLAB 5G Toolbox. The steps used to generate the 5G NR signal are shown in Fig.~\ref{fig:5GCommChain}. The signal includes the transport (uplink shared channel, UL-SCH) and physical channels. The transport block is segmented after the cyclic redundancy check (CRC) addition and low-density parity-check (LDPC) coding is used as forward error correction. The output codewords are 16-QAM modulated. Both data and control plane information is loaded to the time-frequency grid of the 5G signal. Orthogonal frequency-division multiplexing (OFDM) modulation is used with inverse Fast Fourier Transform (IFTT) and Cyclic Prefix (CP) addition operations. The transmit frequency is set to $4$ GHz. The subcarrier spacing is $15$ kHz. The number of resource blocks used in the simulations is $52$. The transmitted waveform is passed through a tapped delay line (TDL) propagation channel model. The delay spread is set to $300\text{x}10^{-9}$ seconds. Additive white Gaussian noise (AWGN) is added to the signal in order to simulate the signal received at the receiver. 

The adversary $A$ collects $1000$ signal samples. Each sample includes sensed signal powers as features and `ACK' or `no ACK' as labels. We  split data samples by half for training and testing of classifier $C_A$. The hyperparameters of $C_A$ are the same as $C_T$ except that the input layer has $400$ neurons. The classifier $C_A$ can correctly detect all ACK instances and most no-ACK instances. The error probability when there is no ACK is $4.8$\%. Once $C_A$ is built, the adversary $A$ can successfully jam all data transmissions or all pilot signals from $T$. Moreover, the unnecessary jamming is minimized, i.e., among all jamming transmissions only $4.48$\% are performed when there is no ACK.

As mentioned earlier, $A$ can jam either $T$'s data transmission or $T$'s pilot signal. In general, $T$'s data transmission period is much longer than $T$'s pilot signal. Thus, jamming of the pilot signal is more energy efficient. For example, if the length of data transmission is nine times of the length of pilot signal and the adversary has the energy to jam data transmissions for $20$\% of time slots, the adversary cannot jam all data transmissions and can only reduce the throughput by $19.08$\%. With the same amount of energy consumption, the adversary can jam all pilot signals and thus reduce the throughput by $100$\%.
 
\subsection{Scenario 2: Adversarial Attack on Signal Authentication in Network Slicing} \label{subsec:Scenario2}

\subsubsection{Attack Setting} \label{subsubsec:Attack2}

\begin{figure}[h]
	\centering
	\fbox{\includegraphics[width=0.95\columnwidth]{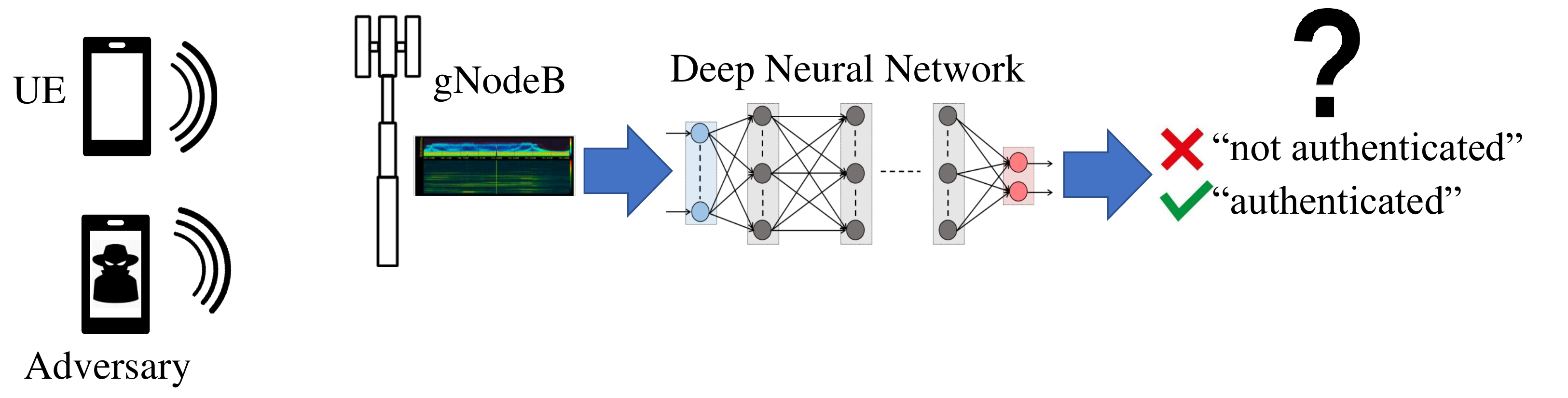}}
	\caption{Scenario 2: 5G signal authentication. 
	\label{fig:Scenario2Step1}}
\end{figure}

The second scenario considers a spoofing attack on the network slicing application. A classifier is trained at the 5G gNodeB to detect 5G UEs by their signals and then provide certain services after authenticating them, as shown in Fig. \ref{fig:Scenario2Step1}. The 5G gNodeB trains the classifier $C_S$ based on the I/Q data that includes both signal power and phase to distinguish signals from a target 5G UE and random noise signals. For that purpose, we use the same deep neural network structure as in scenario 1 except that classifier $C_S$ has the input layer of $400$ neurons.

\begin{figure}[h]
	\centering
	\fbox{\includegraphics[width=0.95\columnwidth]{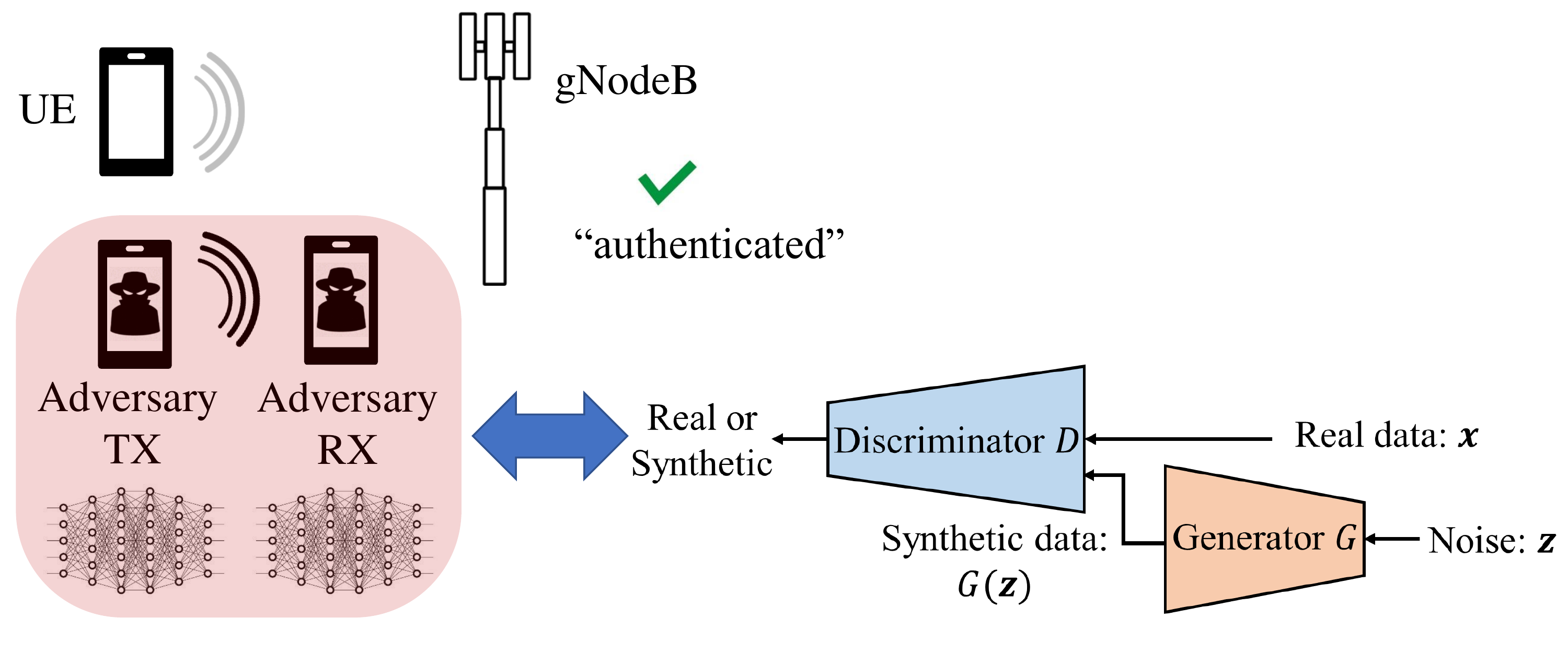}}
	\caption{Scenario 2 - Attack: Adversary trains a GAN over the air to generate spoofing signals and transmits them to infiltrate the 5G signal authentication.
	\label{fig:Scenario2Step2}}
\end{figure}

An adversary $A$ aims to transmit similar signals to gain access to 5G-enabled services. For this purpose, it can sense the spectrum to (i) collect signal samples (I/Q data) and (ii) identify whether such a signal is classified as a target user's signal by either monitoring feedback from the 5G gNodeB (regarding which 5G UE is authenticated) or observing which 5G UE starts communicating to the 5G gNodeB as an authenticated user. Once sufficient 5G signal samples are collected, adversary $A$ can apply the GAN to generate synthetic 5G data as spoofing signals and then transmit them to gain access to 5G-enabled services. In this attack (shown in Fig. \ref{fig:Scenario2Step2}), $A$ consists of a pair of transmitter and receiver. Adversary transmitter $A_T$ trains a neural network to build the generator of the GAN and adversary receiver $A_R$ trains another deep neural network to build the discriminator of the GAN. The only feedback from $A_R$ to $A_T$ during the training is whether the signal is transmitted from the 5G UE or $A_T$. For that purpose, $A_T$ sends a flag along with its signal to $A_R$ to indicate its transmissions and this flag is used to label samples. This attack process is illustrated in Fig. \ref{fig:Scenario2Step2}. Note that this spoofing attack serves the same purpose as an evasion attack, namely fooling the 5G gNodeB into making wrong classification decisions. The only difference is that instead of adding perturbations on top of real transmissions (by jamming the channel as in Scenario 1), the adversary $A$ generates new synthetic signals and transmits them directly to the 5G gNodeB. 

\subsubsection{Simulation Setup and Performance Results} \label{subsubsec:Sims2}

The deep neural network structure for the generator at $A_T$ is given as follows.
\begin{itemize}
	\item The input layer has $400$ neurons.
	\item There are three hidden layers, each a dense layer of $128$ neurons.
	\item The output layer has $400$ neurons.
\end{itemize}

The deep neural network structure for the discriminator at $A_R$ is the same as the generator except that its output layer has two neurons.

For this attack, we set up a scenario, where we vary the noise power with respect to the minimum received 5G signal power at the 5G gNodeB and the corresponding signal-to-noise-ratio (SNR) is denoted by $\gamma$ (measured in dB). The 5G gNodeB builds its classifier $C_S$ by using $1000$ samples, half for training and half for testing. We vary $\gamma$ as $-3$ dB, $0$ dB, and $3$ dB. No matter which $\gamma$ value is used, $C_S$ can always be built perfectly, i.e., there is no error in distinguishing 5G signals from other (randomly generated) signals. Adversary $A$ collects $1000$ samples (5G signals and other signals), applies the GAN to generate synthetic data samples, and transmits them to gain access to services, as shown in Fig.~\ref{fig:Scenario2Step2}. The success probability is shown in Table~14.1. When $\gamma = 3$ dB, the success probability reaches $90\%$. Note that this approach matches all waveform, channel and radio hardware effects of 5G UE's transmissions as expected to be received at the 5G gNodeB. Therefore, this attack performance cannot be achieved by replay attacks that amplify and forward the received signals.

\begin{table}
\caption{Spoofing attack performance.}
\centering
\begin{tabular}{c|c}
5G signal SNR, $\gamma$ (in dB)& Attack success probability \\ \hline \hline
$-3$    & $60.6$\%    \\ \hline
$0$   & $66.6$\%    \\ \hline
$3$   & $90.0$\%
\end{tabular}
\label{}
\end{table}
%\label{table:alpha}

\subsection{Defense against Adversarial Machine Learning in 5G Communications}

\begin{figure} 
	\centering
	\fbox{\includegraphics[width=0.5\columnwidth]{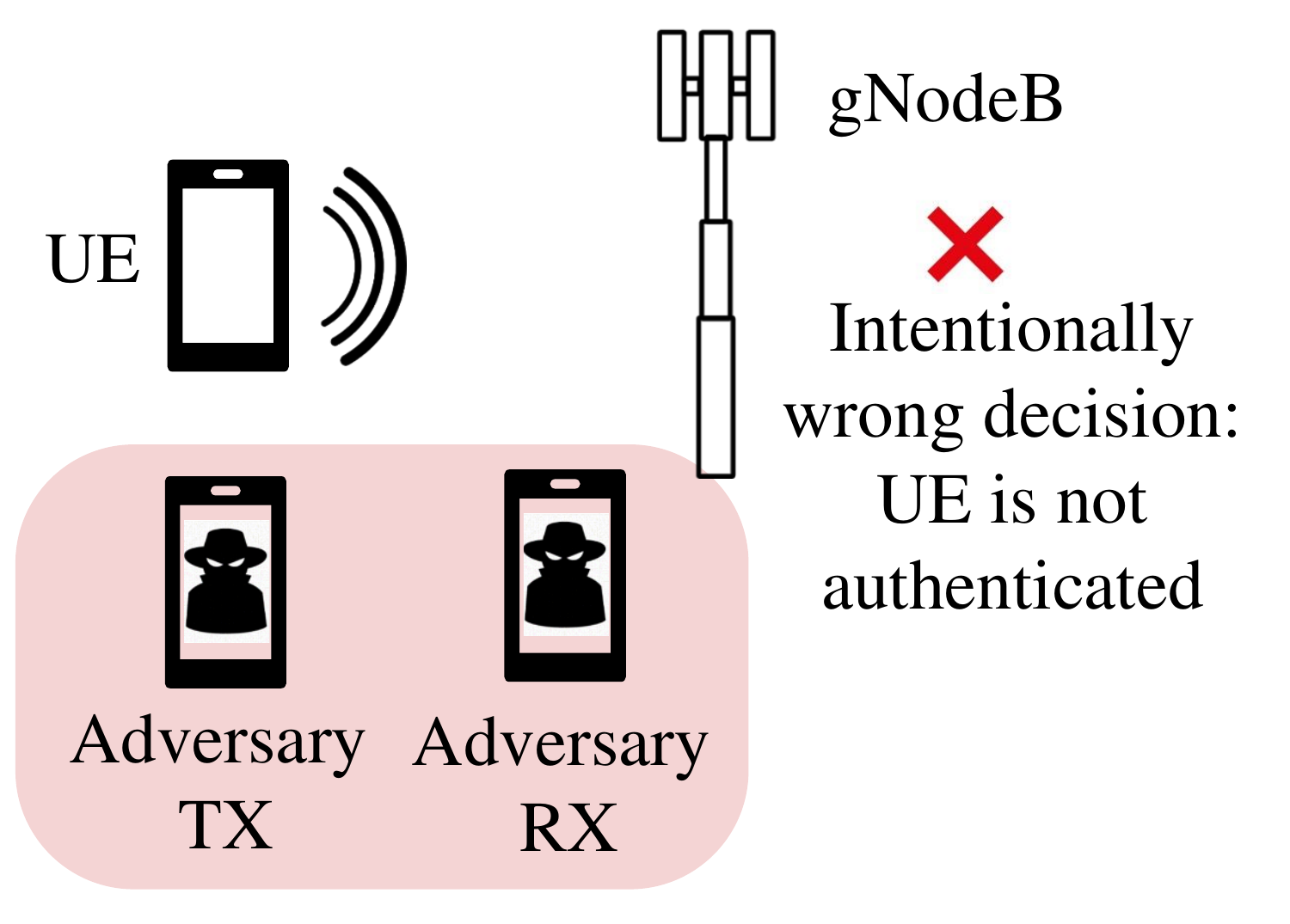}}
	\caption{Scenario 2 - Defense: Controlled errors introduced at 5G gNodeB as part of defense.}
	\label{fig:Scenario2Step3}
\end{figure}

We use the attack scenario 2 (adversarial attack on 5G signal authentication) as an example to discuss the defense. One proactive defense approach at the 5G gNodeB is to introduce selective errors in denying access to a very small number of requests from intended 5G UEs (i.e., false alarm probability is slightly increased). Note that no errors are made in authenticating non-intended users (i.e., misdetection error is not increased). This is not a deterministic approach such that an intended 5G UE that is denied request in one instance can be authenticated in its next access attempt. The controlled errors made by the 5G gNodeB are inputs to the adversary and thus it cannot train a proper GAN to generate synthetic 5G signals, i.e., spoofed signals can be reliably detected and denied access. Fig. \ref{fig:Scenario2Step3} shows the scenario for defense. Given that such defense actions (selective errors) would also decrease the system performance (as very few intended 5G UEs may be denied access over time), the probability of defense actions should be minimized. Thus, the 5G gNodeB selects signal samples that are assigned high score as intended user by its classifier and denies access to their corresponding transmitter. This approach misleads the adversary most, as the uncertainty in determining the decision boundary in its classifier is maximized for a given number of defense actions.

Let $P_d$ denote the ratio of defense actions to all authentication instances. Table~14.2 shows that even a small $P_d$, e.g., $P_d=0.01$, can significantly decrease the attack success probability, where $\gamma$ is fixed as $-3$ dB. On the other hand, there is no need to use a large $P_d$, e.g., larger than $0.05$, since the attack success probability converges to roughly $60$\% quickly.

A similar defense can be applied against scenario 1, where the 5G gNodeB deliberately makes a small number of wrong transmit decisions when accessing the spectrum. Then, the adversary cannot train a proper surrogate model to launch successful attacks on data transmission or spectrum sensing.   

\begin{table}
\caption{Defense against spoofing attack.}
\centering
{
\begin{tabular}{@{}c|c@{}}
Ratio of defense actions, $P_d$ & Attack success probability \\ \hline \hline
$0$   & $90.0$\%    \\ \hline
$0.01$   & $68.2$\%    \\ \hline
$0.02$  & $61.8$\%    \\ \hline
$0.05$   & $59.4$\%    \\ \hline
$0.1$   & $62.0$\%    \\ \hline
$0.2$   & $61.0$\%    
\end{tabular}}
\label{} %{table:alpha}
\end{table}

\section{Conclusion} \label{Conclusion}
The security aspects of machine learning have gained more prominence with the increasing use of machine learning algorithms in various critical applications including wireless communications. We first explained different attack types in adversarial machine learning and corresponding defense methods. Then, we focused on how adversarial machine learning can be used in wireless communications to launch stealthy attacks. We described the challenges associated with designing attacks in wireless domains by accounting for differences from other data domains and unique challenges. Next, we focused on the vulnerabilities of the 5G communication systems due to adversarial machine learning. 

We considered two 5G scenarios. In the first scenario, the adversary learns 5G's deep learning-driven pattern of spectrum sharing with incumbent user such as in the CBRS band and jams the data and control signals to disrupt 5G communications. Results show that the adversary can significantly reduce the throughput of the 5G communications while leaving only a small footprint. In the second scenario, a spoofing attack is performed by the adversary to pass through the deep learning-based physical-layer authentication system at the 5G gNodeB. A GAN is trained to generate the spoofing signal by matching waveform, channel, and radio hardware effects at the receiver. Results show that the attack is successful for a range of SNRs of the 5G signal used during the training. Then, a defense technique is proposed such that controlled errors are made by the 5G system deliberately to fool the adversary into training inaccurate models while minimizing negative effects on its own performance. Novel attacks presented in this paper highlight the impact of adversarial machine learning on wireless communications in the context of 5G and raise the urgent need for defense mechanisms.

\bibliographystyle{plainnat}

\end{document}